\documentclass[useAMS,usenatbib]{mn2e}

\usepackage{graphicx,fleqn}
\usepackage{rotating}
\DeclareGraphicsExtensions{.eps,.ps,.eps.gz,.ps.gz,.eps.Z}
\title[Extreme Properties Of GRB~061007]{Extreme Properties Of GRB~061007: A Highly Energetic Or A Highly Collimated Burst?}

\author[P.~Schady, et al.]{P.~Schady$^{1}$, M.De~Pasquale$^{1}$, M.J.~Page$^{1}$, L.Vetere$^{3}$, S.B.~Pandey$^{1}$, X.Y.~Wang$^{3,5}$, \and J.~Cummings$^{2}$, B.~Zhang$^{4}$, S.~Zane$^{1}$, A.~Breeveld$^{1}$, D.N.~Burrows$^{3}$, N. Gehrels$^{2}$, \and C.~Gronwall$^{3}$, S.~Hunsberger$^{3}$, C.~Markwardt$^{2}$, K.O.~Mason$^{1,6}$, P.~M{\'e}sz{\'a}ros$^{3,7}$, \and J.P.~Norris$^{8,9}$, S.R.~Oates$^{1}$, C.~Pagani$^{3}$, T.S.~Poole$^{1}$, P.W.A.~Roming$^{3}$, P.J.~Smith$^{1}$ \and and D.E.~Vanden Berk$^{3}$\\
$^{1}$ The UCL Mullard Space Science Laboratory, Holmbury St Mary, Dorking, Surrey, RH5 6NT, UK.\\
$^{2}$ NASA/Goddard Space Flight Center, Greenbelt, MD 20771, USA.\\
$^{3}$ Department of Astronomy and Astrophysics, Pennsylvania State University, 525 Davey Laboratory, University Park, PA 16802, USA.\\
$^{4}$ Department of Physics, University of Nevada, Las Vegas, NV 89154, USA.\\
$^{5}$ Department of Astronomy, Nanjing University, Nanjing 210093, China.\\
$^{6}$ The Particle Physics and Astronomy Research Council, Polaris House, North Star Avenue, Swindon, Wiltshire, SN2 1SZ, UK.\\
$^{7}$ Department of Physics, Pennsylvania State University, University Park, PA 16802, USA.\\
$^{8}$ Department of Physics and Astronomy, University of Denver, 2199 S. University Blvd., Denver, CO 80208, USA.\\
$^{9}$ Visiting Scholar, Stanford University, Stanford, CA 94305, USA.
}

\date{Received: }

\begin{document}

\newcommand\eg{e.g. }
\newcommand\ie{i.e. }
\newcommand\swift{{\it Swift}}
\newcommand\invsqrcm{cm$^{-2}$}
\newcommand\nh{$N_{H}$}
\newcommand\av{$A_V$}
\newcommand\Fmax{$F_{\nu,\rm{max}}$}
\newcommand\num{$\nu_m$}
\newcommand\nuc{$\nu_c$}
\def\lesssim{\mathrel{\hbox{\rlap{\hbox{\lower4pt\hbox{$\sim$}}}\hbox{$<$}}}}

\maketitle

\begin{abstract}
GRB~061007 was the brightest gamma-ray burst (GRB) to be detected by \swift\ and was accompanied by an exceptionally luminous afterglow that had a $V$-band magnitude $< 11.1$ at 80~s after the prompt emission. From the start of the \swift\ observations the afterglow decayed as a power law with a slope of $\alpha_X=1.66\pm 0.01$ in the X-ray and $\alpha_{opt}=1.64\pm 0.01$ in the UV/optical, up to the point that it was no longer detected above background in the optical or X-ray bands. The brightness of this GRB and the similarity in the decay rate of the X-ray, optical and $\gamma$-ray emission from 100~s after the trigger distinguish this burst from others and present a challenge to the fireball model. The lack of a cooling or jet break in the afterglow up to $\sim 10^{5}$~s constrains any model that can produce the large luminosity observed in GRB~061007, which we found to require either an excessively large kinetic energy or highly collimated outflow. Analysis of the multi-wavelength spectral and high-resolution temporal data taken with \swift\ suggest an early time jet-break to be a more plausible scenario than a highly energetic GRB. This must have occurred within 80~s of the prompt emission, which places an upper limit on the jet opening angle of $\theta_j=0.8^{\circ}$. Such a highly collimated outflow resolves the energy budget problem presented in a spherical emission model, reducing the isotropic-equivalent energy of this burst to $E_{\gamma}^{corr}=10^{50}$~erg, consistent with other GRBs.
\end{abstract}


\section{Introduction}
During their initial outburst, gamma-ray bursts (GRBs) are the most luminous source of electromagnetic radiation in the Universe, releasing isotropic-equivalent energies of the order of $10^{48}$--$10^{54}$~erg in $\gamma$-rays alone on timescales of $10^{-3}$--$10^3$~s. The favoured mechanism to produce such a large amount of energy on such a short timescale involves accretion onto a newly formed black hole produced either in a merger of two compact objects or from the gravitational core collapse of a massive star. There appear to be two distinct populations of GRBs, distinguished by the duration of their prompt emission and their spectral properties \citep{kmf+93}, whereby long GRBs (prompt emissions lasting $> 2$~s\ \footnotemark[1]) are typically softer than short GRBs ($<2$~s) \citep{sak+06}. The most robust model for long, soft GRBs is the collapsar model \citep{woo93}, in which a GRB results from the energy released during the death of a massive star. Evidence supporting this model includes the large number of long GRBs located in regions of active star-formation \citep[e.g.][]{tbb+04,fls+06}, and the association of several nearby GRBs with supernovae, \eg~GRB~980425/SN1998bw \citep{kfw+98}, GRB~030329/SN2003dh \citep{hsm+03} and GRB~060218/SN2006aj \citep{crc+06}
\footnotetext[1]{The duration of the prompt emission for the BATSE GRB sample showed a bimodal distribution with separation at $\sim 2$~s \citep{kmf+93}. However, in the light of \swift\ the distinction between short and long GRBs has become less clear \citep[e.g. GRB~060614;][]{gnb+06,zzl+07}, and both temporal and spectral properties need to be considered to distinguish between these two populations of GRBs.}

There is evidence to suggest that collimated outflows reduce the amount of energy emitted by a GRB from isotropic emission by $\sim 2$ orders of magnitude, where the beam-corrected energy is clustered at $10^{50}$--$10^{51}$~erg (Frail et al. 2001; Bloom, Frail \& Kulkarni 2003). Bipolar jets are a natural consequence of accretion onto a black hole and are observed in many high-energy astrophysical phenomena (\eg\ AGN, pulsars, micro-quasars), where the ejecta take the easiest exit path, along the rotational axis.

A signature of highly relativistic jets is an achromatic break in the broadband afterglow light curves, caused by the expansion of the radiation beaming angle as the bulk Lorentz factor, $\Gamma$, decreases (Rhoads 1999; Sari, Piran \& Halpern 1999). Once the Lorentz factor has decreased such that $\theta_j= 1/\Gamma$, where $\theta_j$ is the half jet opening angle, the rate of decay increases, resulting in a break in the light curve.

One of the scientific aims of the GRB dedicated mission, \swift\ \citep{gcg+04}, was to pin down the range of collimation present in GRBs. \swift\ is a rapid-response mission with multi-wavelength capabilities designed specifically to study the early time evolution of GRBs and their afterglows. The spacecraft is equipped with three telescopes; the Burst Alert Telescope \citep[BAT; ][]{bbc+05} has a 15--150~keV energy range with a partially coded response up to 350~keV, the X-Ray Telescope \citep[XRT; ][]{bhn+05} which covers the 0.3--10~keV energy range, and the Ultra-Violet and Optical Telescope \citep[UVOT; ][]{rkm+05}, which can observe in six colours and a broadband white band filter, covering the wavelength range from 1600~\AA\ to 6500~\AA.

\swift's slewing capabilities provide GRB afterglow light curves from $\sim 60$~s after the trigger, and well-sampled X-ray and optical light curves have revealed far fewer achromatic breaks than had been expected prior to launch (Ghirlanda, Ghisellini \& Lazzati 2004; O'Brian et al. 2006). This may be an indication that jet breaks occur later than previously believed, at a time when the afterglow is no longer detected above the background level. In the X-ray band this is at around $10^6$~s after the prompt emission. The time of the jet break, $t_j$, increases with the jet opening angle, $\theta_j$, as $t_j\propto \theta_j^{8/3}(n/E_k)^{-1/3}$~erg, for a constant density circumburst medium with density $n$ and GRB kinetic energy $E_k$. Large average jet opening angles, small circumburst densities or large GRB kinetic energies may, therefore, all contribute to delaying the time of the jet-break to when the afterglow is no longer detectable by \swift, although $t_j$ is only weakly dependent on the latter two parameters listed.

GRB~061007 is an example of a burst that showed no evidence of a break in the X-ray or UV/optical afterglow from 80~s after the prompt emission, until it was no longer detected above background, at $\sim10^{6}$~s. Prompt observations by ground based observatories measured the redshift of this burst to be z=1.26 (Osip, Chen \& Prochaska 2006; Jakobsson et al. 2006). At this redshift, GRB~061007 had one of the highest isotropic-equivalent energies ever seen, releasing $\sim 10^{54}$~erg in $\gamma$-radiation alone, with an optical afterglow comparable to that of GRB~990123 \citep{abb+99}, with $V < 11.1$~mag at 71.7~s after the BAT trigger. 

In this paper we describe in detail the $\gamma$-ray, X-ray and optical observations taken by \swift. From the high resolution early time data we constrain the type of environment in which the burst occurred and determine the conditions that distinguish this burst from others. In section~\ref{sec:obs_anlysis} we describe the data reduction and analysis techniques used, followed by an analysis on the possible physical parameters and emission mechanisms that could explain the observations in section~\ref{sec:fireballmod}-\ref{sec:ssc}. In these subsections we investigate both a wind and ISM circumburst environment, as well as spherical-equivalent emission and collimated emission. We also use the afterglow spectral energy distribution (SED) to analysis the properties of the circumburst medium of GRB~061007 in section~\ref{sec:host}. Our results are discussed in section~\ref{sec:model_disc} and in section section~\ref{sec:conc} we summarise our conclusions. Throughout the paper all errors quoted are $1\sigma$ unless otherwise specified and the temporal decay rates, $\alpha$, and spectral indices, $\beta$, are defined such that $F_{t}\propto t^{-\alpha}$ and $F_{\nu}\propto \nu ^{-\beta}$, respectively. Luminosities and the energy release are calculated assuming a standard cosmology model with $H_0 = 70$~km~s$^{-1}$~Mpc$^{-1}$, $\Omega_M=0.3$ and $\Omega_{\lambda}=0.7$.

\section{Observations and Analysis}\label{sec:obs_anlysis}
\subsection{BAT}
The BAT on board \swift\ triggered on GRB~061007 on $7^{th}$ October, 2006 at 10:08:08~UT, resulting in an immediate slew to point the narrow field instruments in the direction of the burst. From here on we shall refer to the time of the trigger as T. The refined BAT position for GRB~061007 is RA = $03^{h}05^{m}11.8^{s}$, Dec = $-50^{\circ} 29\arcmin47.4\arcsec$ (J2000) with an uncertainty (systematic and statistical) of $0.9\arcmin$ at 90\% containment \citep{mbb+06}.

\begin{figure}
\centering
\includegraphics[width=0.5\textwidth]{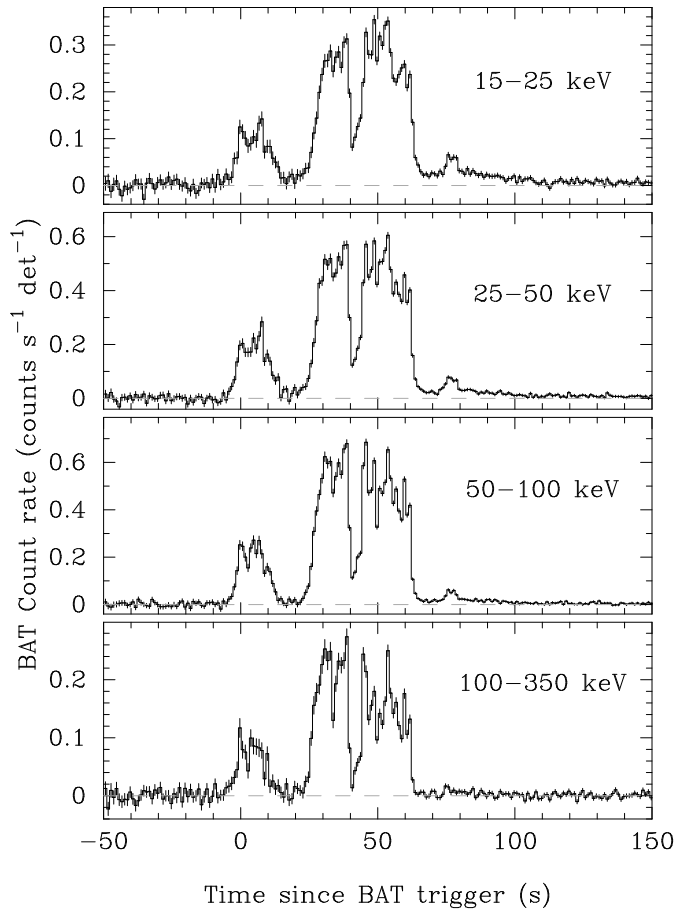}
\caption{Prompt emission light curve of GRB~061007 shown in the four BAT energy bands.}\label{fig:BATlc}
\end{figure}

The BAT light curve consists of three large and distinct peaks and is shown in Fig.~\ref{fig:BATlc}. The time-interval over which 90\% of the 15--150~keV $\gamma$-radiation is emitted was $T_{90}= 75\pm 5$~s, although emission above the background level was detected out to $\sim$T+500~s, which is most pronounced in the soft energy band (top panel of Fig.~\ref{fig:BATlc}). The time lag \citep{nb06} between the peak emission in the 50--100~keV and the 15--25~keV energy bands was $16^{+8}_{-6}$~ms, and between the 100--350~keV and 25--50~keV bands the lag was $21^{+7}_{-5}$~ms.

\subsection{XRT}\label{sec:xrt}
The first XRT observation was a centroiding image taken 80.4~s after the BAT trigger, and a bright uncatalogued source was found at RA = $03^{h}05^{m}19.5^{s}$, Dec = $-50^{\circ}30\arcmin 01.9\arcsec$, with a 90\% containment radius of $3.5\arcsec$ \citep{vpb+06}.

At 86.6~s after the BAT trigger the XRT began taking data in Window Timing mode (WT), in which data are stored with a 1.7~ms time resolution and 1-dimensional imaging. At T+2000~s the count rate fell to below $\sim 1$ count~s$^{-1}$ and the automated sequence switched to Photon Counting (PC) mode, in which the time resolution is 2.5~ms and both imaging and spectroscopic information are available. The X-ray afterglow was detected for $10^{6}$~s after the BAT trigger before it fell below the detection threshold. During the first 2200~s of observations the count rate of the burst was high enough to cause pile-up in both  WT and PC mode data. To account for this effect, the WT data were extracted in a rectangular 40$\times$20-pixel region with a 9$\times$20-pixel region excluded from its centre. The size of the exclusion region was determined following the procedure illustrated in \citet{rcc+06}. The standard correction for pile-up was applied to the PC data, as described by \citet{vgb+06}. Due to a large amount of pile-up in the first 30~s of WT observations, these data were not used in our spectral analysis.

\begin{figure*}
\centering
\includegraphics[width=0.9\textwidth]{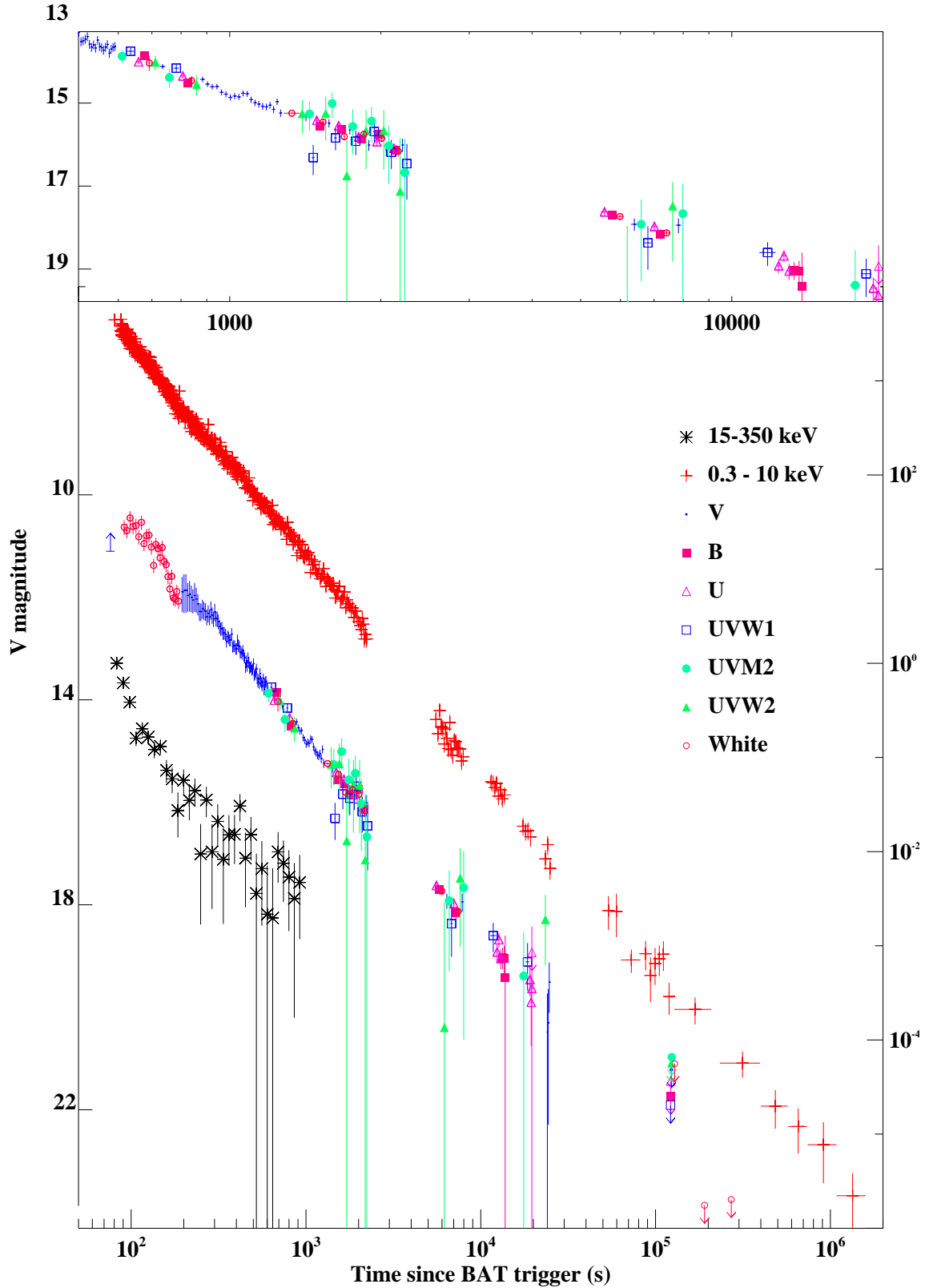}
\caption{GRB afterglow light curve shown in all six UVOT lenticular filters, in the X-ray (0.3--10~keV) and in the $\gamma$-ray (15--350~keV). All light curves have been shifted vertically with respect to the $V$ band filter for clarity. The $B$, $U$, $UVW1$, $UVM2$, $UVW2$ and white filter have all been normalised to the $V$ band light curve. The left hand axis corresponds to the $V$-band light curve, and the count rate axis on the right applies to the X-ray and $\gamma$-ray light curves. The best fit decay indices to the combined UVOT light curve, the X-ray light curve and the BAT light curve are $\alpha_{UVOT}=1.64\pm 0.01$, $\alpha_{XRT}=1.66\pm 0.01$ and $\alpha_{BAT}=1.54\pm 0.12$, respectively. Error bars are smaller than the symbol size if not visible.}\label{fig:comblc}
\end{figure*}

The 0.3--10~keV light curve is well fit by a power law with decay index $\alpha=1.66\pm 0.01$ from T+80~s until at least T+$10^6$~s, when it reached the detectability threshold of XRT (see Fig.~\ref{fig:comblc}).  We also considered the presence of a break in the light curve, resulting from either the edge of the jet coming into the line of sight or the migration of the cooling frequency, \nuc, through the observing energy bands. The cooling frequency defines the frequency above which electrons no longer lose a significant fraction of their energy to radiation. The cooling frequency is expected to evolve as $t^{-1/2}$ for an ISM-like circumburst environment (Sari, Piran \& Narayan 1998), and as $t^{1/2}$ for a wind-like circumburst environment \citep{cl00}, and as it migrates through the observing energy band the drop in radiation at $\nu _{obs} > \nu _c$ produces a break in the light curve.

To determine the $3\sigma$ lower limit on the time at which a jet or cooling break could have occurred, we applied a broken power law fit modelled on the cooling break, for which $\alpha _2=\alpha _1 +0.25$ \citep{spn98}, and on a jet break, where $\alpha_2 =\alpha_1 + 0.75$ \citep{rho99}, and in each case determined the earliest time that the break could occur within the observing time interval without deviating the goodness of fit by more than $3\sigma$ from the best-fit power law model. Our analysis indicates that a cooling break could not have occurred earlier than $10^5$~s after the BAT trigger to 99\% confidence, and any jet break after the onset of the afterglow observations must have happened at $t_j > 2.65\times 10^5$~s.

A power law fit to the early WT spectrum ($<$T+600~s), where the Galactic column density is fixed at $N_H=2.13\times 10^{20}$~\invsqrcm\ \citep{dl90}, gives a best-fit host galaxy column density of $N_H(\rm(WT))=(5.3\pm 0.2)\times 10^{21}$\ \invsqrcm\ at a redshift of $z=1.26$ for a spectral index $\beta_X(\rm(WT)) = 0.96\pm 0.01$. Spectral analysis of the PC data during the time interval T+5000~s to T+25,000~s gives a best fit host galaxy column density $N_H(\rm{PC})=(5.1\pm 0.6)\times 10^{21}$~\invsqrcm, and spectral index $\beta_X(\rm(PC))=0.92\pm 0.5$, consistent with the WT data, and therefore indicating that there is no spectral evolution in the X-ray data.

\subsection{UVOT}\label{sec:uvot}
The UVOT began observing GRB~061007 71.7~s after the BAT trigger, at which point it took a 9~s settling exposure in the $V$-filter. This was followed by a 100~s exposure in the white light filter ($\lambda\sim 1600$--$6500$~\AA) and a 400~s exposure in the $V$-band. After this the automated sequence rotated twice through the UVOT filters taking a series of short exposures ($V, UVM2, UVW1, U, B$, white, $UVW2$) (10~s for $B$ and white filters and 20~s for the rest). The central wavelengths of the optical and UV filters are $5460$~\AA, $4370$~\AA\, $3460$~\AA, $2600$~\AA, $2260$~\AA\ and $2030$~\AA\ for the $V$, $B$, $U$, $UVW1$, $UVM2$ and $UVW2$, respectively. A further 400~s $V$-band exposure and 100~s white exposure were taken, followed by a further series of rapid rotations around the filter wheel up to 2260~s after the trigger. With the exception of the second 100~s white band exposure, all observations up to this time were taken in event mode, which has a time resolution of 11~ms, and after this image mode exposures were taken. The refined position of the afterglow is RA = $03^{h}05^{m}19.6^{s}$, Dec = $-50^\circ 30\arcmin02.4\arcsec$ to a certainty of $0.5\arcsec$ (J2000). The Galactic extinction along this line of sight is $E(B-V)=0.021$ (Schlegel, Finkbeiner \& Davis 1998).

For all short exposures (10~s or 20~s) and exposures taken later than T+2260~s photometric measurements were obtained from the UVOT data using a circular source extraction region with a $6\arcsec$ radius for the $V$, $B$ and $U$ optical filters and a $12\arcsec$ radius for the three UVOT ultra-violet filters to remain compatible with the current effective area and zeropoint calibrations\footnotemark[1].
\footnotetext[1]{http://heasarc.gsfc.nasa.gov/docs/heasarc/caldb/swift/docs/uvot/}
The background was measured in a source-free region near to the target using an extraction radius of $20\arcsec$. 

The first $V$-band settling exposure and first white-band exposure were affected by coincidence loss to a degree that is outside the photometrically calibrated range of UVOT; for a $V$-band settling exposure complete saturation within a standard $6\arcsec$ aperture implies that this afterglow is brighter than $V = 11.1$ for the whole of the 9~s exposure.

During the first white band exposure the source is so bright that relative photometry can be obtained using a very large aperture, which includes the wings of the point spread function, where coincidence loss is not a problem. The exposure was divided into 5~s bins, and source counts were obtained from an extraction region of radius $\sim 26\arcsec$. To calibrate this photometry, observations of the white dwarf standards GD50 and HZ2 were taken with UVOT and the white filter on October 18$^{th}$ 2006. From these data we determined the relationship between observed count rate within the $26\arcsec$ aperture and corrected count rate within a standard $6\arcsec$ UVOT source aperture, which we assume to be linear. To apply this calibration to GRB~061007 we apply constant offset to this relation to account for the different background count rate between the GRB~061007 observations and the white dwarf standard count rate observations. This offset was chosen to match the mean white-band count rate in the saturated exposure to the white filter light curve extrapolated from later times; we caution that calibration of this first white exposure is crude and thus the apparent steepening of the light curve towards the end of the exposure may not be real. However, the overall behaviour of a steadily decreasing count rate over the course of the exposure is not in doubt. For the first 400~s $V$-band exposure a higher resolution light curve was created by splitting it into 4~s bins, and the second into 20~s bins, consistent with the majority of the exposures during that time.

In order to get the best measurement of the optical temporal decay we created a single UV/optical light curve from all the UVOT filters. To do this the light curve in each filter was individually fitted to find the corresponding normalisation, and this was then used to renormalised each light curve to the $V$-band. The combined UV and optical light curve, from T+85~s up to T+$2.5\times 10^4$~s, is best fit with a power law decay with temporal index $\alpha_{opt}=1.64\pm 0.01$. The combined light curve shows no apparent colour evolution, illustrated in Fig.~\ref{fig:comblc}.

The combined UVOT light curve is shown next to the 0.3--10~keV and 15--350~~keV light curve in Fig.~\ref{fig:comblc}, and a close up of the UVOT light curve from T+1000~s to T+30,000~s is shown in the inset. The log-log scale in this figure shows the decaying light curve in the 15-350~keV energy band from 80--500~s much better than the linear scale in Fig~\ref{fig:BATlc}. The magnitude along the left axis corresponds to the $V$-band, and the count rate along the right axis applies to the X-ray and $\gamma$-ray light curves.

\subsection{Observations By Other Facilities}
The prompt emission from GRB~061007 also triggered Konus-Wind and Suzaku, and the afterglow was monitored by ground based telescopes, which carried out both spectroscopic and photometric observations. GRB~061007 was measured to be at z=1.26 \citep{ocp06,jft+06}, for which the rest-frame isotropic energy release and maximum luminosity in the 20~keV--10~MeV is $E_{iso}\sim 1.0\times 10^{54}$~erg, and $L_{iso}\sim 1.8\times 10^{53}$~erg~s$^{-1}$ \citep{gam+06}. The optical afterglow was also detected by ROTSE at 26.4~s after the BAT trigger at a magnitude of 13.6~mag in an unfiltered exposure \citep{rr06}, and the Faulkes Telescope South measured an $R$-band magnitude of $R=10.15\pm 0.34$ 142~s after the trigger \citep{bmg+06}, consistent with the UVOT observations and indicative of a significant brightening during the first 90~s. The \swift\ observations from T+80~s show that by this time the afterglow had entered a constant power law decay (see Fig.~\ref{fig:comblc}). Radio observations were also performed 1--1.24~days after the prompt emission, which resulted in no detection of a radio source, and a flux measurement of $1\pm36~\mu$Jy at 19~GHz \citep{vr06}.

\subsection{Multi-Wavelength Light Curve Analysis}\label{sec:multiLC}
At $\sim T+100$~s, after the last flare in the BAT light curve (Fig.~\ref{fig:BATlc}), the BAT light curve appears to decay smoothly and monotonically out to $\sim$T+500~s, after which the flux becomes indistinguishable from the background in the BAT energy range. A power law fit to the BAT light curve from T+100~s to T+500~s provides an acceptable fit ($\chi ^2 = 32$ for 29 dof) with a best-fit decay index of $\alpha_{BAT} = 1.61\pm 0.14$, which is consistent with the decay observed in both the UV/optical and X-ray light curves. Furthermore, a power law spectral fit to this last part of the BAT light curve, beyond $\sim$T+100~s, gives a best fit energy spectral index $\beta_{BAT}(>T+100~\rm{s}) = 0.80\pm 0.09$, in fairly good agreement with the spectral indices determined for the X-ray and optical afterglows. This supports a scenario in which the $\gamma$-emission at this epoch is generated by the same radiation mechanism as the X-ray and UV/optical afterglow.

\begin{figure*}
\centering
\includegraphics[width=0.7\textwidth]{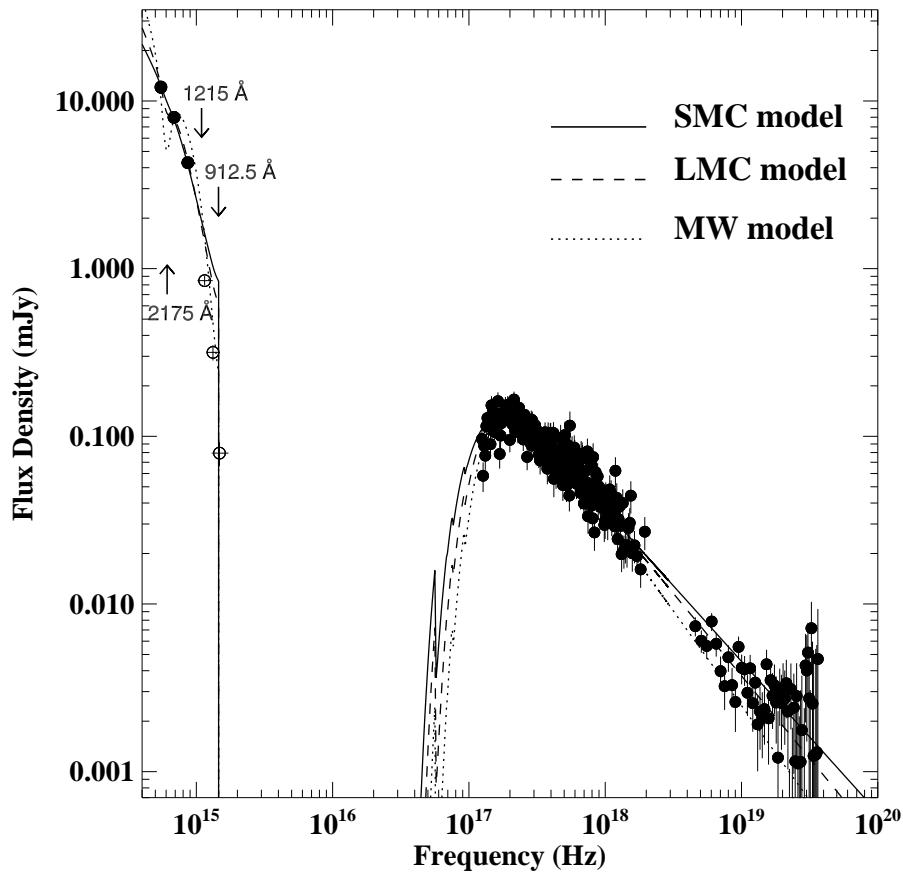}
\caption{The combined $\gamma$-ray, X-ray and UV/optical spectral energy distribution of GRB~061007 at T+600~s (see text) with best-fit models for each corresponding dust extinction curve shown; SMC (dashed), LMC (solid) and Galactic (dotted). Open circles are data points at wavelength $\lambda < 1215$~\AA\ in the rest frame, and therefore not used in the spectral fitting. The rest-frame positions of the Galactic absorption feature at 2175~\AA\, the beginning of the Lyman-$\alpha$ forest at 1215~\AA\, and the Lyman-$\alpha$ break at 912.5~\AA\ are indicated.}\label{fig:SED}
\end{figure*}

\subsection{The Spectral Energy Distribution}\label{sec:sed}
Using the full range of data available with \swift\ an SED of the afterglow was produced using optical, X-ray and $\gamma$-ray data, which were normalised to an instantaneous epoch at 600~s after the BAT trigger. At this time there are data available from all three instruments, and good colour information from UVOT. For each of the UVOT lenticular filters the tool {\sc uvot2pha} (v1.1) was used to produce spectral files compatible with {\sc xspec} (v12.2.1) from the exposures taken closest to T+600~s. The count rate in each band was then normalised to the estimated value at T+600~s determined from the best-fit models to the light curves. The XRT spectrum was produced from WT mode data in the 0.5--10~keV energy range, omitting the first 30~s of data, which suffered from pile-up. This was scaled to the corresponding count rate at T+600~s. For the BAT spectrum, data in the 15--150~keV energy range were taken from the time interval T+100~s -- T+500~s, during which the BAT light curve decayed at a rate consistent with X-ray and UV/optical afterglows and had a similar spectrum. This spectrum was normalised to T+600~s in the same way as the XRT spectrum. The SED is shown in Fig.~\ref{fig:SED}.

The lack of any break in the X-ray and UV/optical light curves up to $10^5$~s places the cooling frequency either above the X-ray band or below the optical at T+600~s. Both energy bands will, therefore, have the same spectral index, and the SED was, thus, fit with a power law emission spectrum, and two dust and gas components were included to model the Galactic and host galaxy photoelectric absorption and dust extinction. The column density and reddening in the first absorption system were fixed at the Galactic values (see sections~\ref{sec:xrt} and \ref{sec:uvot}). The second photoelectric absorption system was set to the redshift of the GRB and the neutral hydrogen column density in the host galaxy was determined assuming solar abundances. The dependence of dust extinction on wavelength in the GRB host galaxy was modelled on three extinction laws taken from observations of the Milky Way (MW), the Large Magellanic Cloud (LMC) and the Small Magellanic Cloud (SMC). The greatest differences observed in these extinction laws are in the amount of far UV (FUV) extinction, which is greatest in the SMC and least in the MW, and in the strength of the $2175$~\AA\ absorption feature, which is most prominent in the MW and negligible in the SMC. To model these extinction laws the parameterisations from \citet{pei92} were used, where $R_V=A_V/E(B-V)=3.08, 2.93$, and 3.16 for the Galactic, SMC and LMC extinction laws respectively.

An important difference between the Magellanic Clouds and the Milky Way is the average metallicity of these galaxies, which is larger in the case of the Milky Way. The value that we determine in our spectral analysis for \nh\ is an equivalent neutral hydrogen column density that results from the amount of soft X-ray absorption in the spectrum, which will, therefore, depend on the abundances assumed in the spectral fit. The SMC and LMC have a metallicity that is $\sim 1/8$ and $\sim 1/3$ that of the Milky Way \citep{pei92}, and this will result in an underestimation of the neutral hydrogen column density by a factor of $\sim 8$ and $\sim 3$, respectively, if solar abundances are assumed. This is taken into account when we discuss the results from our spectral analysis in section~\ref{sec:host}.

\begin{table*}
\begin{center}
\caption{Power law spectral fit results to the spectral energy distribution of GRB~061007}\label{tab:SEDpowfit}
\begin{tabular}{@{}llllll}
\hline
Spectral Model & Host galaxy & Rest-Frame & $\beta_{o,X}$& $\chi ^2$ & Null-hypothesis\\
 & \nh$^{ 1}$ ($10^{21}$~\invsqrcm) & visual extinction, \av & & (dof) & probability\\
\hline\hline
SMC & $3.66^{+0.29}_{-0.28}$ & $0.39\pm 0.01$ & $0.90\pm 0.005$ & 398 (308) & $4\times 10^{-4}$\\
LMC & $5.30^{+0.33}_{-0.32}$ & $0.66\pm 0.02$ & $0.98\pm 0.007$ & 347 (308) & 0.063\\
MW & $7.62^{+0.81}_{-0.77}$ & $1.13\pm 0.10$ & $1.13^{+0.04}_{-0.03}$ & 1042 (308) & 0.000\\
\hline
\end{tabular}
\end{center}
$^{1}${\footnotesize \ Equivalent hydrogen column density assuming solar abundances at the GRB host galaxy.}
\end{table*}

At a redshift of $z=1.26$ the beginning of the Lyman-$\alpha$ forest is redshifted to an observer-frame wavelength of $\sim 2580$~\AA, which falls squarely on the $UVW1$ filter, the reddest of the UV filters. The unknown equivalent width of Lyman-$\alpha$ absorption from the host galaxy of the GRB and the variation of the Ly-$\alpha$ forest along different lines of sight makes it difficult to model this additional source of absorption in the SED. We, therefore, exclude the three data points affected by Lyman-$\alpha$ from our spectral analysis. The results from our spectral analysis are summarised in Table~\ref{tab:SEDpowfit}. The fitted value of the spectral index depends, for the most part, on the amount of optical/UV extinction, which in turn is affected by the dust extinction model used. Similarly, the amount of soft X-ray absorption fitted to the model will affect the best-fit spectral index.

In all three cases an absorption and extinction system is measured at the GRB host galaxy, which is smallest in an SMC model and largest in a MW model. The lack of any evolution in $N_H$ or $\beta$ in the X-ray data (see section~\ref{sec:xrt}), indicates that $E_p$ has already moved through the X-ray band, and, therefore, the neutral hydrogen column density measured is indicative of the amount of X-ray absorption at the GRB host galaxy rather than an artifact of evolution in the spectral index. The optical, X-ray and $\gamma$-ray spectrum is best fit by a model that assumes the GRB host galaxy to have an LMC extinction law for the GRB host galaxy, which gives a $\chi^2 = 347$ for 308 degrees of freedom (dof) with a null-hypothesis probability of $p=0.063$. Models with an SMC and MW extinction law at the host galaxy were rejected with 99.96\% confidence and 100\% confidence, respectively.

\section{Discussion}\label{sec:disc}
GRB~061007 is one of the brightest bursts to be observed by \swift, with an isotropic energy released in $\gamma$-rays that is second only to GRB~050904 \citep{tac+05} and an optical flux comparable to that observed in GRB~990123 \citep{abb+99}. The similarity of the afterglow of GRB~061007 in the flux decay of $\gamma$-ray, X-ray and UV/optical bands, and the lack of features in X-ray and optical light curves distinguishes this burst from others.

Optical afterglows typically decay as power laws, although usually with decay indices $\sim 1.1$ (Zeh, Klose \& Kann 2006), and in the X-ray, the GRB light curve is usually characterised by several discrete segments, with a canonical shape that is made up of three, and sometimes four, power law segments with different decay indices \citep{nkg+06,zfd+06}. The temporal behaviour of GRB~061007 is, therefore, not typical of other \swift\ GRBs, of which only $\sim 7\%$ observed promptly by the XRT (within the first few minutes of the BAT trigger) show a constant decay with no additional features such as flaring \citep{woo+06} (e.g. GRB~051117B, GRB~060403). When only considering long GRBs, the fraction of smooth decaying, featureless X-ray afterglows is even more infrequent, amounting to only $\sim 4$\% of the long GRB population. 

According to the fireball model, the temporal and spectral indices of the afterglow depend on the properties of both the GRB ejecta and the circumburst environment, which determine the location of the characteristic synchrotron frequency, \num\, and the cooling frequency, \nuc, with respect to the observing band \citep{zlp+07}. At any one time the location of these frequencies are dependent on the energy distribution of electrons in the circumburst environment defined by the energy index $p$, the density and density profile of the circumburst medium, the amount of kinetic energy in the outflow, and on the fraction of energy in electrons and in the magnetic field, $\epsilon_e$ and $\epsilon_B$, respectively.  $\epsilon_e$ and $\epsilon_B$ are typically assumed to remain constant with time, and therefore the evolution of \num\ and \nuc, and thus the afterglow decay and spectral indices, depend on the density  and density profile of the circumburst medium and on the amount of kinetic energy released by the GRB.

The lack of a break in the light curves provide constraints on the spectral regime that the X-ray and UV/optical afterglow are in, and on the frequency of \num\ and \nuc, neither of which could have crossed the observing window from T+80~s to T+$10^5$~s. From the white filter we determine that around half the counts detected in the white band observations correspond to the $V$ and $B$ filter. The lack of a break in the first white filter exposure, therefore, indicates that at T+80~s, \num\ was no bluer than the $B$-band filter. This corresponds to an upper limit on the characteristic synchrotron frequency $\nu_m < 7.5\times 10^{14}$~Hz. From the UV/optical data we also determine that at T+80~s the flux density in the $V$-band, corresponding to $\nu = 5.5\times 10^{14}$~Hz, is $F_{\nu} = 730$~mJy, after correcting for absorption at the host galaxy. 

With these constraints in place we investigate the various conditions that can explain the temporal and spectral properties of GRB~061007, all of which can be categorised into two parent models that differ fundamentally in the effect of collimation on the observations. In the first model we assume no jet break to have occurred before T+$2.65\times 10^5$~s, based on the constraints we obtained in section~\ref{sec:xrt}, and the observed emission is, therefore, indistinguishable from isotropic emission. Secondly, we investigate a model in which the jet break occurs before our first X-ray and UV/optical observations at T+80~s such that the assumption of isotropic emission no longer applies.

\subsection{Spherical Expansion Model}\label{sec:fireballmod}
The simplest model to consider is one in which the observed afterglow emission follows the flux decay and spectral evolution as would be expected from synchrotron radiation emitted from an isotropic outflow. As already mentioned, one of the important factors that determines the evolution of the defining synchrotron frequencies is the density profile in the surrounding circumburst material, and we therefore consider both a wind-like circumburst environment, such as would be expected in the vicinity of a massive star, and a constant density profile, such as in the ISM. For simplicity, we assume the density profile of the circumburst medium to have the form $\rho\propto r^{-s}$, where $s=0$ and $s=2$ corresponds to an ISM and wind-like profile, respectively.

The relation between $\alpha$ and $\beta$ for synchrotron emission at different intervals in the synchrotron spectrum are defined by a set of closure relations, both for a constant density medium \citep{spn98} and a wind density profile \citep{cl00}. Assuming spherically equivalent emission, the standard closure relations do not fully satisfy the observed spectral and temporal indices of GRB~061007 for either an ISM or a wind-like circumburst medium, where $\alpha_{o,X}=1.65\pm 0.01$ and $\beta_{o,X}=0.99\pm 0.02$. This could be the result of additional components unaccounted for in the closure relations, or an oversimplification of the assumptions made when outlying the closure relations \eg\ the density profile may not be distributed as $r^{-s}$.

For both a wind and an ISM circumburst density profile our observations are in closest agreement with a model in which the observed synchrotron emission is in the slow cooling phase ($\nu_m < \nu_c$), where the UV/optical and X-ray bands are in the same slow-cooling regime ($\nu_m<\nu<\nu_c$). The decay index is $\Delta\alpha\sim 0.17$ steeper than expected for a constant circumburst density profile, and requires a fairly soft electron distribution index $p\sim 3$. In the case of a wind circumburst environment the decay index is shallower than predicted by this model, by a difference $\Delta\alpha\sim0.33$. It is possible that the density profile in the environment surrounding GRB~061007 lies somewhere between the two scenarios considered here, where $0<s<2$. Alternatively, the shallowness of the decay index for a wind case could be explained if energy injection is involved. This would require a late-time source of energy into the forward shock that would slow down the cooling rate of the excited electrons and, thus, slow down the afterglow decay. Either way, the lack of a cooling break for the duration of \swift\ observations out to $10^6$~s requires consideration, and the conditions necessary to maintain \nuc\ above or below the observing frequency range need to be investigated. 

The evolution of \nuc\ with time is dependent on the density profile of the circumburst medium, whereby for $s > \frac{4}{3}$ \nuc\ will increase with time, whereas for $s < \frac{4}{3}$ it will decrease \citep{pan06}. The conditions needed for the cooling frequency to not have crossed the observing window, therefore, differ for an ISM and wind circumburst density profile. For an ISM medium the cooling frequency has to be maintained above the X-ray band for up to T+$10^5$~s, whereas in a wind scenario the cooling frequency needs to be above the X-ray band at the start of the X-ray observations, at T+80~s. Either way the cooling frequency must lie above the peak X-ray frequency out to at least T+$10^5$~s for a cooling break not to cover in the X-ray observations (see section~\ref{sec:xrt}). This corresponds to the requirement that $\nu_c > 6\times 10^{17}$~Hz at T+$10^5$~s. In the next section we investigate the consistency of a spherical emission model with the data for both an ISM and wind medium where $\nu_m < \nu < \nu_c$.

\subsubsection{ISM Case}\label{sec:ismmod}
For an ISM-like circumburst environment, the temporal and spectral flux decay indices are related as $\alpha = 3\beta/2$. For a value of $\beta_X = 0.99\pm 0.02$, we expect $\alpha=1.49\pm 0.03$, which is shallower than the observed decay rate of $\alpha= 1.65\pm 0.01$. However, this deviation from the relation is small, and could be accommodated if the density profile is slightly different from the uniform density we have assumed. In this regime $\beta = (p-1)/2$, where $p$ is the electron energy index. Thus, for $\beta_X=0.99\pm 0.02$ the electron energy index is in the range $2.94 < p < 3.02$. We therefore take $p = 3.0$. These relations hold as long as the characteristic synchrotron frequency, $\nu_m$, lies below the optical band, \ie\ $\nu_m < 7.5\times 10^{14}$~Hz, at T+80~s, as discussed previously. 

For a constant density medium in which synchrotron radiation is the dominating emission mechanism, the characteristic synchrotron frequency, the cooling frequency and the peak density flux are given by the following equations
\begin{equation}
\nu_m  =  3.3\times 10^{15}\Bigl(\frac{p-2}{p-1}\Bigr)^2(1+z)^{1/2}\epsilon^{1/2}_{B}\label{eq:num}\\
\epsilon^2_{e} E^{1/2}_{k,52}t^{-3/2}_d~{\rm Hz}
\end{equation}
\begin{equation}
\nu_c  =  6.3\times 10^{12}(1+z)^{-1/2}(1+Y)^{-2}\epsilon^{-3/2}_{B}\\
 E^{-1/2}_{k,52}n^{-1}t^{-1/2}_d~{\rm Hz} \label{eq:nuc}
\end{equation}
\begin{equation}
F_{\nu,{\rm max}}=16(1+z)D_{28}^{-2}\epsilon^{1/2}_{B}E_{k,52}n^{1/2}~{\rm mJy}\label{eq:Fmax}
\end{equation}
 \citep{zlp+07}, where $\epsilon_B$ is the fraction of energy in the circumburst magnetic field, $\epsilon_{e}$ is the fraction of energy in electrons, $E_{k,52}$ is the kinetic energy in the GRB in units of $10^{52}$, $t_d$ is the time since the prompt emission in units of days, $n$ is the density of the surrounding circumburst medium, $Y$ is the Compton parameter, and $D_{28}$ is the luminosity distance in units of $10^{28}$~cm. Typical values for the fraction of energy in the cicumburst magnetic field and in the electrons are $\epsilon_B = 0.01$ and $\epsilon_e=0.1$, respectively. 

For a spectral index $\beta = 1$, $\nu F_{\nu}$ is constant for $\nu_m < \nu < \nu_c$. At T+80~s $F_{\nu} = 730$~mJy at $\nu = 5.5\times 10^{14}$~Hz, and therefore $\nu_mF_{\nu,{\rm max}} = 730\times (5.5\times 10^{14})$. By using this together with Eq.~\ref{eq:num} and \ref{eq:Fmax} we obtain:
\begin{equation}\label{eq:density}
n =\frac{3.4\times 10^{-6}}{\epsilon^2_{B}\epsilon^4_{e}E^3_{k,52}}
\end{equation}

In Eq.~\ref{eq:nuc} we apply the condition that at $10^5$~s after the prompt emission \nuc\ must still be greater than $6\times 10^{17}$~Hz, and substituting in Eq.~\ref{eq:density} gives
\begin{eqnarray}\label{equal1}
(1+Y)^{-2}\epsilon^{1/2}_{B}\epsilon_e^4E^{5/2}_{k,52} > 0.52
\end{eqnarray}

Considering now the characteristic synchrotron frequency, we know that \num $< 7.5\times 10^{14}$~Hz, and this reduces Eq.~\ref{eq:num} to
\begin{eqnarray}\label{equal2}
\epsilon^{1/2}_{B}\epsilon^2_{e}E^{1/2}_{k,52} < 1.7\times 10^{-5},
\end{eqnarray}
This, and the inequality given in Eq.~\ref{equal1} provide an upper limit to the value of $\epsilon_{B}$ in terms of $E_{k,52}^3$, such that
\begin{eqnarray}\label{eq:epsilonBlims}
\epsilon_{B} < 3.1\times 10^{-19}(1+Y)^{-4}E_{k,52}^3
\end{eqnarray}
The above equation bounds us to extreme values of $E_{k,52}$ and $\epsilon_{B}$. For $E_{k,52} = 10^3$, $\epsilon_B<3.1\times 10^{-10}(1+Y)^{-4}$, which is an unrealistically low value of $\epsilon_B$. Even a value as small as $\epsilon_B=10^{-6}$ requires a huge amount of kinetic energy to meet the condition that $E_{k,52} > 1.5\times 10^4(1+Y)^{4/3}$~erg. The amount of kinetic energy assumed is an isotropic-equivalent value which would, thus, be smaller if it were corrected for collimation. A $3\sigma$ lower limit of $t_j = 2.65\times 10^5$~s can be imposed as the earliest time that a jet break could have occurred. $t_j$ is the time of the jet break measured from the onset of the afterglow, which we take to be equal to T. A kinetic energy $E_{k,52}=1.5\times 10^4(1+Y)^{4/3}$~erg and $\epsilon_B=10^{-6}$ would require $\epsilon_e=0.012(1+Y)^{-1/3}$ and $n=49(1+Y)^{-8/3}$~cm$^{-3}$ for the conditions given above to be satisfied. From this we estimate the jet opening angle using
\begin{eqnarray}\label{eq:thetaj}
\theta_j = 0.2\Bigl(\frac{t_{j,d}}{1+z}\Bigr)^{3/8}\Bigl(\frac{n}{E_{k,52}}\Bigr)^{1/8},
\end{eqnarray}
\citep{pk00}, where $t_{j,d}$ is in units of days. The equation for the IC parameter is as follows:
\begin{eqnarray}
Y = [-1 +(1+4\eta_1\eta_2 \epsilon_e/\epsilon_B)^{1/2}]/2
\end{eqnarray}
\citep{zlp+07}, where $\eta_1=\min[1,(\nu_c/\nu_m)^{(2-p)/2}]$ \citep{se01}, and $\eta_2\ll 1$ accounts for the Klein-Nishina correction. 
For the physical parameters assumed, the IC parameter, $Y$, is near negligible $(Y^6(1+Y)<2.2\times 10^{-11})$, and can therefore by approximated to $Y=0$. In this case, a jet opening angle for GRB~061007 would need to be $\theta_j > 6.3^{\circ}$ to produce a jet break later than T+$2.65\times 10^5$~s. This would provide a beam-corrected kinetic energy of $E_k^{corr}> 9\times 10^{53}$~erg,  which is half a solar mass and three orders of magnitude greater than usual bursts \citep{fw01}. The energetics required by this model are unreasonably large and, therefore, an unlikely solution to explain the properties of GRB~061007. 

\subsubsection{Wind Case}\label{sec:windmod}
In the slow cooling regime where $\nu_m < \nu < \nu_c$, the spectral and temporal indices for a wind-like circumburst environment are related such that $\alpha = (3\beta + 1)/2$, which can be rejected by the observations at more than $10~\sigma$ confidence. In order for this model to be compatible with the data an additional source of energy is required to slow down the decay rate of the afterglow. A continual source of energy injection could come from either the central engine itself, or late time refreshed shocks that slow down the decay of the afterglow. The shallow decay phase observed in many XRT light curves is interpreted to be the result of energy injection, and although this phase typically lasts for $10^4$~s, there are also examples of GRBs with evidence of energy injection for prolonged periods of time, and with very smooth decay slopes (\eg\ Romano et al. 2006; Huang, Cheng \& Gao 2006). Taking this into account, an explanation whereby continual energy injection maintains the afterglow and slows down its decay over a long period of time is not unreasonable. For the purpose of this analysis the source of this additional energy is not considered and the energy injection is simply quantified by the expression $L\propto t^{-q}$, where L is luminosity in units of erg~s$^{-1}$ \citep{rm98}. A wind circumburst environment may then be compatible with our observations if  $\nu_m < \nu < \nu_c$ and $q=0.66\pm 0.02$ and $p= 2.98\pm 0.02$. Using these best fit parameters and the known constraints provided by the data it is possible to determine the parameter space of the GRB kinetic energy and the circumburst medium microphysical parameters that could reproduce the observations. 

For a wind environment the characteristic synchrotron frequency, \num, and flux at this frequency are given by
\begin{equation}\label{eq:windnum}
\begin{array}{lll}
\nu_m & = & 4.0\times 10^{15}\Bigl(\frac{p-2}{p-1}\Bigr)^2(p-0.69)(1+z)^{1/2}\\
& & \epsilon^{1/2}_B\epsilon_e^2E_{K,52}^{1/2}t_d^{-3/2}~\rm{Hz}
\end{array}
\end{equation}
\begin{equation}\label{eq:windFnum}
F_{\nu,\rm{max}} = 77(p+0.12)(1+z)^{3/2}D^{-2}_{28}\epsilon_B^{1/2}E_{K,52}^{1/2}A_*t^{-1/2}~\rm{mJy}
\end{equation}
where $A_* = A/(5\times 10^{11})$~g~cm$^{-1}$ is the density scaling such that $\rho=Ar^{-2}$ \citep{cl00}. As before the flux at $\nu=5.5\times 10^{14}$~Hz is $F_{\nu} = 730$~mJy at T+80~s, and since $\nu F_{\nu}=\nu_mF_{\nu,\rm{max}}$, we have
\begin{equation}
A_* = \frac{9.16\times 10^{-7}}{\epsilon_B\epsilon_e^2E_{K,52}}\label{eq:windA}
\end{equation}
The dependencies on the cooling frequency also vary from Eq.~\ref{eq:nuc} for a wind medium, where it is now defined by
\begin{eqnarray}\label{eq:windnuc}
\begin{array}{lll}
\nu_c & = & 4.4\times 10^{10}(3.45-p)e^{0.45p}(1+z)^{-3/2}\\
& & (1+Y)^{-2}\epsilon_B^{-3/2}E_{K,52}^{1/2}A^{-2}_*t_d^{1/2}~\rm{Hz},
\end{array}
\end{eqnarray}
which has been modified from \citet{cl00} to include the Compton parameter, $Y$. A new expression, therefore, needs to be defined regarding the conditions on the cooling frequency. Since \nuc\ increases with time, the only useful constraint is a lower limit on \nuc\ at early times. This constraint can be determined through spectral modelling of the afterglow SED at an epoch of T+300~s, which is the weighted mid-time of the time interval over which the BAT spectrum shown in Fig.~\ref{fig:SED} was accumulated. This SED was fit with a broken power law with a spectral break above the BAT energy range, and the change in spectral index at the break was fixed to $\Delta\beta=0.5$ to correspond to a cooling break. Taking the $\chi^2$ from this fit into account, the \nuc\ upper limit was then determined by shifting the spectral break to lower energies until the goodness of fit deviated by $\Delta\chi^2=9$ ($3\sigma$ confidence for one interesting parameter). This gave an upper limit on the cooling frequency of $\nu_c > 1.2\times 10^{19}$~Hz at T+300~s. Applying this, Eq.~\ref{eq:windnuc} combined with Eq.~\ref{eq:windA} reduces to
\begin{eqnarray}\label{windequal}
(1+Y)^{-2}\epsilon_B^{1/2}\epsilon_e^4E_{K,52}^{5/2} > 0.0073
\end{eqnarray}
As for an ISM circumburst environment, $\nu_m<7.5\times 10^{14}$, which if applied to Eq.~\ref{eq:windnum} gives
\begin{eqnarray}
\epsilon_B^{1/2}\epsilon_e^2 E_{K,52}^{1/2}<6.26\times 10^{-6},
\end{eqnarray}
and combining this expression with Eq.~\ref{windequal} gives
\begin{equation}\label{eq:windepsilonBlims}
\epsilon_B<2.88\times 10^{-17}(1+Y)^{-4}E_{K,52}^3
\end{equation}
For $\epsilon_B = 10^{-6}$ the kinetic energy in the forward shock needs to be $E_{K,52} > 3\times 10^3(1+Y)^4$~erg, which when taking into account the possible jet collimation reduces to $E_K^{corr} > 1.8\times 10^{53}(1+Y)^4$~erg. However, it is also necessary to include the energy injected into the fireball, which increases the total energy in the forward shock with time as $E_{K}\propto t^{1-q}$. For $q=0.66$ the total energy in the forward shock at T+$10^5$~s increases by an order of magnitude to  $E_K^{corr} > 1.8\times 10^{54}(1+Y)^4$~erg, which is similar to the amount of kinetic energy required for an ISM-like circumburst environment. A wind environment, therefore, faces the same energy budget problem as present for an ISM environment

In the spherical emission model discussed above the primary assumptions made in both the ISM and wind scenarios are that both the X-ray and UV/optical afterglow observed are produced by the same radiation mechanism, and that the decay rate is equivalent to spherical emission. The difficulty in accommodating the observations with a standard fireball model indicate that in the case of GRB~061007 either synchrotron emission is not the dominant radiation mechanism throughout the observing energy range, or that the GRB outflow is collimated whereby the observed emission is no longer equivalent to isotropic emission. With this in mind we investigate how additional radiation mechanisms and various collimation effects would alter the observed afterglow from that produced by synchrotron emission from an isotropic outflow.

\subsection{Highly Collimated Models}
The large amount of energy required to power a model with isotropic emission leads us to investigate a scenario in which the emission from the GRB is highly collimated, and we once again assume the circumburst environment to be of constant density. A very narrow jet opening angle could reduce the energy required to explain our observations by several orders of magnitude, where $t_j$ must be prior to our first afterglow observations in order to account for the lack of a break in the X-ray and UV/optical light curves at later times. If we assume the afterglow begins no earlier than the time of trigger, this corresponds to a time $t_j < 80$~s. Within the premise that the outflow is highly collimated we consider two cases in which the cooling frequency is either above or below the observing frequencies. In these two regimes $p$ and $\beta$ are related by $\beta=(p-1)/2$ and $\beta=p/2$, respectively, corresponding to $p\sim 3.0$ or $p\sim 2.0$. After the jet break the flux decays as $F_t\propto t^{-p}$ \citep{sph99}, where $p$ is either 2 or 3, which is steeper than the observed decay rate of $\alpha\sim 1.65$. However, this does not account for continual energy injection, which would reduce the afterglow decay rate.

\subsubsection{\num $< \nu <$ \nuc}\label{sec:numid}
For a laterally expanding jet with constant density circumburst medium and \num $< \nu <$ \nuc, the relation between the temporal decay of the GRB afterglow, $\alpha$, and the spectral index, $\beta$, post jet-break is given by
\begin{equation}
\alpha = (1+2\beta) - \frac{2}{3}(1-q)(\beta+2),
\end{equation}
\citep{pmg+06} where $q$ is the energy injection parameter such that $L=L_0(t/t_0)^{-q}$ \citep{zm01}, and since we take the time of our jet break to be at $t_j=80$~s, $t_0=t_j$, $L=L_0(t/t_j)^{-q}$. For our observed temporal and spectral indices, $q=0.3$ and $p=3$, as before. Prior to the jet break
Eq.~\ref{eq:num}-\ref{eq:Fmax} still apply, but in order take into account the effect of energy injection, the kinetic energy requires an additional factor of $(t/t_j)^{(1-q)}$. The time of jet break and the jet opening angle are related as $\theta_j\propto t_j^{3/8}$. To keep $\theta_j$ as large as our observations will allow we take $t_j=80$~s. At T+80~s the additional factor of $(t/t_j)^{(1-q)}$ is unity and there is no change in Eq.~\ref{eq:num}-\ref{eq:Fmax}. Thus at T+80~s, Eq.~\ref{eq:density} and Eq.~\ref{equal2} stand. For Eq.~\ref{eq:nuc} the constraint that $\nu_c > 6\times 10^{17}$ at T+$10^5$~s means that the evolution of \nuc\ post jet-break has to be taken into account. If $\nu_{c_j}$ is the frequency of $\nu_c$ at the time of the jet break, T+80~s,
\begin{equation}\label{eq:nucstart}
\nu_{c_j} = 4.05\times 10^{19}(1+Y)^{-2}\epsilon_B^{1/2}\epsilon_e^4E_{k,52}^{5/2}
\end{equation}
However, we are concerned with where $\nu_c$ lies at T+$10^5$~s. After a jet break $\nu_c\propto \gamma^{-4}t^{-2}$, where $\gamma\propto t^{-(2+q)/6}$ is the Lorentz factor of the electrons in the circumburst environment \citep{pmg+06}. The cooling frequency after T+80~s is given by
\begin{equation}\label{eq:postjetnuc}
\nu_c = \nu_{c_j}\Bigl(\frac{\gamma}{\gamma_j}\Bigr)^{-4}\Bigl(\frac{t}{t_j}\Bigr)^{-2}=\nu_{c_j}\Bigl(\frac{t}{t_j}\Bigr)^{\frac{2}{3}(q-1)},
\end{equation}
where $\gamma_j$ is the electron Lorentz factor at T+80~s. Using the condition that $\nu_c > 6\times 10^{17}$~Hz at T+$10^5$~s, for $q=0.3$ Eq.~\ref{eq:postjetnuc} reduces to 
\begin{equation}\label{eq:evnuc}
(1+Y)^{-2}\epsilon_B^{1/2}\epsilon_e^{4}E_{k,52}^{5/2} > 0.41
\end{equation}
In the same way that we determined the inequality given in Eq.~\ref{eq:epsilonBlims}, we combine the above equation with Eq.~\ref{equal2} and determine that
\begin{equation}\label{eq:epsilonBlims2}
\epsilon_B < 5.0\times 10^{-19} (1+Y)^{-4} E_{k,52}^3
\end{equation}
For $E_{k,52}=5\times 10^4$~erg, $\epsilon_B<6.2\times 10^{-5}(1+Y)^{-4}$, and if we take this upper limit as the value of $\epsilon_B$, $\epsilon_e = 3.1\times 10^{-3}(1+Y)^{1/2}$. Although the assumed kinetic energy is large, this corresponds to the isotropic-equivalent value and the beam-corrected kinetic energy is a few orders of magnitude smaller than this. For $p=3$, the inverse Compton parameter $Y$ is proportional to $[\eta(\epsilon_e/\epsilon_B)]^{1/2}$, where $\eta$=min[$1,(\nu_m/\nu_c)^{1/2}$], and for these values of $\epsilon_e$ and $\epsilon_B$, $Y<0.3$. For values, of $E_{k,52}=5\times 10^4$, $\epsilon_B=6.2\times 10^{-5}$, and $\epsilon_e=3.1\times 10^{-3}$ the particle density in the circumburst environment is $n=0.08$~cm$^{-3}$, which, using Eq.~\ref{eq:thetaj} gives an opening angle of $\theta_j=2.0\times 10^{-3}$~rad. This corresponds to $\theta_j=0.11^{\circ}$, which gives a beam-corrected kinetic energy $E_{k}^{corr} = 10^{51}$~erg and $\gamma$-ray energy $E_{\gamma}^{corr}=2\times 10^{48}$~erg. This would imply an efficiency of $\eta\approx 0.002$, which is two orders of magnitude smaller than is typically inferred \citep{zlp+07}. The jet opening angle that we estimate assumes the hydrodynamics as described by \citet{pk00}, and variations on the hydrodynamics, such as that discussed in \citet{sph99}, result in a different coefficient in Eq.~\ref{eq:thetaj}. However, these differences are of order unity, and do not affect the validity of the model discussed.

This model is, therefore, able to satisfy our observations without the requirement of an excessively large kinetic energy. However, the opening angle is a factor of at least 30 narrower than previously observed \citep{bfk03} and the inferred $\gamma$-ray energy is extremely low for a long GRB, prompting us to consider other options.

\subsubsection{$\nu > \nu_c$}\label{sec:nuup}
Keeping in place the condition that $t_j< 80$~s, we investigate a model in which the X-ray and UV/optical bands are above \num\ and \nuc. In this model our observational constraints on the values of \num, \nuc\ and \Fmax\ only apply at T+80~s. Since we assume a jet break to occur at $t_j=80$~s, the factor of $(t/t_j)^{(1-q)}$ in the kinetic energy is unity, and $q$ is therefore not of importance in our calculations. For \nuc\ above the observing band, $\beta=p/2$, and therefore $p=2$, in which case the dependence on the minimum Lorentz factor of the shock accelerated electrons $\gamma_m$ changes from
\begin{equation}
\gamma_m = \frac{(p-2)}{(p-1)}\epsilon_e\frac{\rm{m}_p}{\rm{m}_e}\Gamma
\end{equation}
 to
\begin{equation}
\gamma_m = \rm{ln}\Bigl(\frac{\varepsilon_m}{\varepsilon_M}\Bigr)\epsilon_e\frac{m_p}{m_e}\Gamma,
\end{equation}
where $\varepsilon_m$ and $\varepsilon_M$ are the minimum and maximum energy of the shock accelerated electrons, and $\Gamma$ is the bulk Lorentz factor in the shocked medium \citep{pz06}. $\nu_m\propto \gamma_m$, and therefore the expression for \num\ given in Eq.~\ref{eq:num} has to be modified such that $[(p-2)/(p-1)]^2$ is replaced by [ln$(\varepsilon_m/\varepsilon_M)]^2$. The value of ln$(\varepsilon_M/\varepsilon_m)$ is not well determined due to the unknown magnetic field in the upstream of the GRB shock, although it generally ranges between 5 and 10 \citep{lw06}. We therefore take ln$(\varepsilon_M/\varepsilon_m) = 7$, such that Eq.~\ref{eq:num} becomes
\begin{equation}\label{eq:modnum}
\nu_m  =  6.7\times 10^{13}~\rm{Hz}(1+z)^{1/2}\epsilon^{1/2}_{B}
\epsilon^2_{e} E^{1/2}_{k,52}t^{-3/2}_d
\end{equation}
We assume the same jet break time as before ($t_j=80$~s), whereby Eq.~\ref{eq:nuc} and \ref{eq:modnum} reduce to
\begin{equation}\label{eq:rednuc}
\epsilon_B^{-3/2}E_{k,52}^{-1/2}n^{-1}(1+Y)^{-2} < 5.45
\end{equation}
and
\begin{equation}\label{eq:rednum}
\epsilon_B^{1/2}\epsilon_e^2E_{k,52}^{1/2}< 2.1\times 10^{-4}
\end{equation}
In the slow cooling regime, where $\nu_m<\nu_c$,
\begin{eqnarray}\label{eq:xjetFnu}
\begin{array}{lll}
F_\nu & = & F_{\nu,\rm{max}}\Bigl(\frac{\nu_c}{\nu_m}\Bigr)^{-(p-2)/2}\Bigl(\frac{\nu}{\nu_c}\Bigr)^{-p/2}\\
  & & F_{\nu,\rm{max}}\nu_c^{1/2}\nu_m^{(p-1)/2}\nu^{-p/2}
\end{array}
\end{eqnarray}
By taking the measured value of $F_{\nu} = 730$ at $\nu~=~5.5\times 10^{14}$~Hz, and using Eq.~\ref{eq:nuc}, \ref{eq:Fmax} and \ref{eq:modnum}, the expression for $F_{\nu}$ reduces to
\begin{eqnarray}\label{eq:redFmax}
\epsilon_eE_{k,52}(1+Y)^{-1} = 3.65
\end{eqnarray}
From these equalities we determine that a kinetic energy of $E_{k,52} = 10^3$~erg would require $\epsilon_e=3.7\times 10^{-3}(1+Y)$, $\epsilon_B < 0.25(1+Y)^{-4}$ and $n>0.05(1+Y)^4$. For a GRB local environment with a particle density of $n=10^4$~cm$^{-3}$, which is a typical density for a molecular cloud \citep{rp02}, the jet opening angle would be $\theta_{j}=0.8^{\circ}$, a factor of a few smaller than that determined for some other GRBs \citep[e.g. GRB~980519; ][]{naa+99}. This gives a beam-corrected kinetic energy of $E_{k,52}^{corr}=9.7\times 10^{50}$~erg.

This model provides an alternative scenario in which a highly collimated jet is the cause for the large flux observed in GRB~061007, but where a larger jet opening angle and more standard beam-corrected $\gamma$-ray can satisfy our observations, with respect to the previous model.

\subsection{Synchrotron Self-Compton Emission}\label{sec:ssc}
For completeness we look into a scenario in which synchrotron self-Compton (SSC) emission contributes to the observed emission. In the case where the X-ray emission is dominated by SSC radiation, and the UV/optical emission is produced by synchrotron radiation, both observed bands would need to be in the fast-cooling regime, with $\nu_{opt} > \nu_c > \nu_m$ and $\nu_X > \nu_c^{IC} > \nu_m^{IC}$, in order for the decay slope and spectral index in the X-ray and UV/optical emission to be consistent. In this case the spectral index is given by $\beta = p/2$, which would give a value of $p = 2.0$. However, the decay of the optical afterglow is expected to be $-(2-3p)/4$, which is inconsistent with our observations of $\alpha = 1.65\pm 0.01$.

\subsection{The Circumburst Medium}\label{sec:host}
From our spectral analysis we determine the dust extinction in the circumburst medium to be most consistent with a wavelength dependence similar to that of the LMC. This is consistent with previous studies of GRB environments (\eg\ Fruchter et al. 2006, Kann, Klose \& Zeh 2006; Schady et al. 2006), which also find GRB hosts to be irregular galaxies, and is also in agreement with the collapsar model, which requires a sub-solar metallicity progenitor star, with an upper limit of $Z_{\odot}\lesssim 0.3$ (\eg\ Hirschi, Meynet \& Maeder 2005; Woosley \& Heger 2006).

The best-fit parameters for the X-ray column density and dust extinction local to the GRB are \nh\ $=(5.3\pm 0.3)\times 10^{21}$~\invsqrcm\ and \av\ $=0.66\pm 0.02$, and this gives a gas-to-dust ratio of \nh/\av\  $=(8.0\pm 0.5)\times 10^{21}$~\invsqrcm. However, \av\ was determined assuming an LMC dust extinction law, whereas \nh\ was determined assuming solar abundances, which are unlikely to be correct for an irregular, Magellanic-type galaxy. Correcting the X-ray column measurement to a metallicity of 1/3 solar, which is appropriate for an ISM similar to that of the LMC, we find that the gas-to-dust ratio is a factor of more than three larger than that measured in the LMC. We note that an environment's metallicity affects the amount of dust present as well as the equivalent \nh\ measured. Our assumptions on the metallicity, therefore, do not change the \av/\nh\ ratio as we measure it, since to first order both \av\ and equivalent \nh\ scale directly with metallicity. It is, therefore, valid to compare the dust-to-gas ratio in different environments derived this way, independent of their metallicity. Although not consistent with the LMC, the dust-to-gas ratio measured in the local environment of GRB~061007 is consistent with the average gas-to-dust measured in the local environment of a sample of previous \swift\ GRBs \citep{smp+07}, which was $\langle N_H/A_V\rangle = (6.7\pm 1.1)\times 10^{21}$~\invsqrcm.

The lack of evolution observed in the soft X-ray absorption later than $\sim 80$~s after the prompt emission suggests that the burst must have completely ionised the gas in its immediate environment within this time scale. The absorbing medium that we observe must, therefore, be far enough from the burst that it was not completely ionised in the first few 100~s of seconds after the initial explosion. The probability of there being an absorption system in the line of sight from an unrelated, intervening galaxy that is of sufficient optical depth to contribute a measurable amount of soft X-ray absorption is very small \citep{oj97,crc+06}, leading us to conclude that the absorbing material measured in the spectrum of GRB~061007 is local to the burst. From the results in \citet{pl02} we estimate the absorbing medium to lie at a few tens of parsecs from the GRB, and the large value of \nh\ suggests that GRB~061007 was embedded in a dense, star-forming region, or giant molecular cloud, in common with many other GRBs (e.g. Dai \& Lu 1999; Wang, Dai \& Lu 2000; Reichart \& Price 2002).

\subsection{Comparing Between Models}\label{sec:model_disc}
It is interesting to note that with $E_{\gamma,iso} = 10^{54}$~erg in the 10~keV--20~MeV energy range, and a rest-frame peak energy of $E_{p} = 902$~keV, GRB~061007 satisfies the Amati relation, given by $E_{p}\propto E_{\gamma,iso}^{1/2}$ \citep{aft+02}. It also satisfies a correlation between the peak isotropic luminosity, $L_{p,iso}$, and spectral lag in the prompt emission, $\tau_{lag}$, whereby $L_{p,iso}\propto \tau_{lag}^{-1.1}$. Both these correlations are satisfied by a large fraction of long GRBs \citep[e.g.][]{ama06,gnb+06}, suggesting that GRB~061007 has the same class of progenitor as other long GRBs.

The consistency observed in the spectral and temporal afterglow behaviour in the X-ray and UV/optical bands, and in the late time $\gamma$-ray band, implies that the emission observed comes from the same spectral segment. If we take the afterglow to be in the slow cooling regime and assume a jet break to occur at $t_j> 2.65\times 10^5$~s a large amount of kinetic energy, of at least $E_k^{corr}=9\times 10^{53}$~erg is required for a constant density circumburst medium, or $E_k^{corr} > 6.0\times 10^{52}$~erg for a wind-like circumburst medium. The implausibly large amount of energy required by both these scenario makes them unsatisfactory models to explain the properties of GRB~061007.

We also investigate a second scenario in which a highly collimated outflow produces a jet break before T+80~s. This requires a jet angle no larger than $0.8^{\circ}$, which satisfies the observational requirements with a lower energy budget and more reasonable set of parameters. The upper limit on the jet opening angle of $0.8^{\circ}$ is a factor of 4 smaller than that previously determined in GRBs (Berger, Kulkarni \& Frail 2003), although still physically possible. Furthermore, the need to observe the jet break in order to measure $\theta_j$ introduce selection effects, and the smaller and larger end of the current jet opening angle distribution will, thus, be underestimated. For the very smallest of jet opening angles to be measured, very early time data of the afterglow is needed, preferentially covering several energy bands.

For a jet break earlier than T+80~s our observations are valid for a spectral regime in which $\nu_m<\nu<\nu_c$, and $\nu_c<\nu$, where in the latter a larger jet opening angle is obtained by assuming a density $n=10^4$~cm$^{-3}$, consistent with that observed in molecular clouds. The dependence of the circumstellar particle density on $\epsilon_B$, $\epsilon_e$ and $E_{k,52}$ in the former of these models does not allow for such high densities, for which we estimate $n=0.08$~cm$^{-3}$. Our spectral analysis on the afterglow resulted in a large column density, which supports a scenario in which the GRB is embedded within a dense molecular cloud or active star-forming region. This therefore favours a model in which $\nu_c<\nu$, where the number density is inferred to be larger.

In the case of a uniform jet, where the energy and Lorentz factor are constant across the jet, the afterglow after a jet break should decay as $t^{-p}$. In both the post jet-break models that we consider the predicted decay slope is steeper than the observed value of $\alpha = 1.65\pm 0.01$. A possible explanation is continual energy injection, which would maintain the energy in the afterglow and slow down the rate of decay. The requirement for steady energy injection over the entire light curve is of some concern, but seems plausible on the basis of many other examples of energy injection in \swift\ afterglow observations. Evidence of prolonged energy injection in the form of plateaus in X-ray light curves is observed in a large fraction of \swift\ GRBs, although typically up to $10^3$-$10^4$~s after the prompt emission. In the case of GRB~061007 a steady energy injection is needed for at least an order of magnitude longer than this, during which no break is observed in the light curve. This is not to say that the central engine needs to be active for this length of time, but simply that it is active for the duration of the prompt emission and ejects shells with a large distribution in the Lorentz factor. A range in $\Gamma$ can provide a source of energy to the afterglow for a long time after the central engine has stopped through ongoing refreshed shocks \citep[e.g.][]{rm98,sm00}. Nevertheless, an energy injection mechanism that remains extremely constant for such a long time requires specific conditions to be in place, such as a central engine luminosity that varies smoothly with time in the case of a long-lasting engine, or a smooth distribution of Lorentz factors in the case of refreshed shocks \citep{zfd+06}.

Alternatively, the slow decay rate of the afterglow in the context of post jet-break could also be the result of a jet-edge effect in which there is no sideways expansion. In such a case the afterglow decay index would steepen by 3/4 when $\Gamma=1/\theta_j$, and would continue to decay at this rate due to the lack of lateral expansion \citep{mr99}. In such a scenario there is no need for energy injection, although additional conditions would be required to prevent the jet from expanding further, such as a high magnetic field within the jet.

A possibly more natural form of non-laterally expanding jet is a structured jet or outflow with an energy distribution that is a decreasing function of angular distance from the jet axis, $\theta_0$, and a narrow, high-energy core at the axis. A jet break is, then, observed when the edge of the uniform inner core becomes visible to the observer, and it is, therefore, the viewing angle rather than the jet half-opening angle that determines the time of the jet break. In order for a jet break to occur before T+80~s the viewing angle needs to be close to the jet axis, which is not that unreasonable given the brightness of GRB~061007. At the time of jet break the smoother distribution in energy, $\epsilon$, compared to a uniform jet results in a smaller change in the decay slope that will be shallower than the 3/4 increase in decay index observed in a uniform jet break. For a power law distribution of $E$, where $E\propto\theta^{-k}$, the smaller the value of $k$ the shallower the afterglow light curve will be after the jet-break. In the case of $\nu_m<\nu_{obs}<\nu_c$, where $p=3$, a decay rate after a jet break with index $\alpha=1.65$ will occur in structured jet with an $E$ distribution index $k=0.3$, and for $\nu_c<\nu_{obs}$ we require $k=1.4$ \citep{pan05}.

A consequence of a very early jet break time is the faster evolution of $\Gamma$ into a non-relativistic phase, which will produce an additional break in the light curve as the decay rate slows down. For a uniform jet with no energy injection, this break would occur at at a time $t_{\rm{nr}}=t_j\theta_j^{-2}$ \citep{lw00}. For $t_j=80$~s and $\theta_j<0.8^{\circ}$, the jet would, therefore, not enter the non-relativistic phase until more than $4\times 10^5$~s after the prompt emission, which is not in conflict with our observations, and in the presence of energy injection or for structured jets $t_{nr}$ would only get larger.

In the context of post-jet break emission, a decay slope of $\alpha=1.65$ can, therefore, be accounted for by a non-laterally expanding jet as well as continual energy injection through refreshed shocks. We, therefore, find an early jet break before T+80~s can satisfy the conditions imposed by our observations without requiring a large energy budget and with a reasonable set of physical parameters.

\section{Conclusions}\label{sec:conc}
In this paper we have presented multi-wavelength spectral and temporal data taken with \swift\ of GRB~061007, an very bright GRB with unusually smooth panchromatic decay. Using these data we explored the circumstances that could produce such a bright burst, and so determined the conditions that distinguish GRB~061007 from other, more standard GRBs. On many levels we find GRB 061007 to be typical of other GRBs, suggesting that the same progenitor model for most long GRBs is also applicable to GRB~061007. The surrounding dust has properties most consistent with an LMC-like environment, and the circumburst material is likely to be dense, suggesting that it is embedded in a large molecular cloud or a region of active star formation. Furthermore, its consistency with both the Amati and timelag-luminosity relations, which apply to a large fraction of long GRBs, indicating that the source of energy in GRB~061007 is no different than that of most other long GRBs. The enormous brightness of GRB~061007 is, therefore, unlikely to be due to a different type of progenitor, but the result of extreme values for a few defining characteristics.

Two inherently different models can satisfy our observations of GRB~061007, where the main difference lies in the time of the jet break; either before or after our observations. Both require certain extreme conditions to account for the brightness of this burst, either in the form of a large energy budget or in the collimation of ejecta into a very narrow jet opening angle. However, given the brightness of GRB~061007, it should not be a surprise that some of the defining properties are non-standard. The large amount of kinetic energy required in a late-time jet break model lead us to favour a model in which the GRB~061007 is produced by the concentration of energy into a highly narrow jet $<0.8^{\circ}$; a condition that can more easily be met than the production of large amounts of energy.

The brightness of GRB~061007 across the electromagnetic spectrum makes it ideal to explore the conditions surrounding the production of GRBs, and its uniqueness provides further insight into the range in properties that GRBs have. The small opening angle that we determine for GRB~061007 has important implications on the true range in jet opening angles, and this could provide a clue as to the cause for the large fraction of \swift\ GRBs with no clear signs of jet breaks.

\section*{ACKNOWLEDGEMENTS}
We gratefully acknowledge the contributions of all members of the \swift\ team. SZ thanks PPARC for support through an Advanced fellowship.

\end{document}